\documentclass[preprint,showpacs,preprintnumbers,amssymb]{revtex4}
\RequirePackage{amsmath2000}

\begin{document}
\title{Granularity-induced gapless superconductivity in NbN films: evidence of thermal phase fluctuations}
\author{G. Lamura$^{1,\dag}$, J.-C. Vill\'{e}gier$^2$, A. Gauzzi$^{3,*}$, J. Le Cochec$^{1,\S}$,
J.-Y. Laval$^{1}$, B. Pla\c{c}ais$^4$, N. Hadacek$^2$, and J.
Bok$^1$}

\affiliation {$^{1}$ LPS-ESPCI, 10, rue Vauquelin, 75231 Paris,
France.
\\$^{2}$CEA-Grenoble, SPSMS-LCP, 17, rue des Martyrs, 38054 Grenoble, France.
\\$^{3}$MASPEC-CNR, Area delle Scienze, 43010 Parma, Italy.
\\$^{4}$LPMC-ENS, UMR 8551, 24, rue Lhomond, 75231 Paris,
France.}

\date{\today}
\begin{abstract}
Using a single coil mutual inductance technique, we measure
the low temperature dependence of the magnetic penetration depth
in superconducting NbN films prepared with similar critical
temperatures $\approx$16 K but with different microstructures.
Only (100) epitaxial and weakly granular (100) textured films
display the characteristic exponential dependence of conventional
BCS \textit{s}-wave superconductors. More granular (111) textured
films exhibit a linear dependence, indicating a gapless state in
spite of the \textit{s}-wave gap. This result is quantitatively
explained by a model of thermal phase fluctuations favored by the
granular structure.
\end{abstract}

\leftline\\

\maketitle

\newpage
Distinctive property of simple superconducting materials like
metals and alloys is the absence of thermally activated electronic
excitations at low temperature $T \ll T_c$, where $T_c$ is the
critical temperature. This property is consistent with a model of
superconducting gap $\Delta_\textbf{k}$ with \textit{s}-wave
symmetry, as predicted by the BCS theory. In contrast, in complex
materials like heavy fermions \cite{gro}, cuprates \cite{har} or
rhutenates \cite{bon}, thermal excitations of arbitrarily small
energies are experimentally observed. This gapless property has
been commonly explained by an excitation spectrum with nodal
points or lines, characteristic of $\Delta_\textbf{k}$ with
\textit{p}- or \textit{d}-symmetry respectively. According to
theoretical considerations, non-\textit{s} symmetries are favored
by non-phonon mediated pairing mechanisms like antiferromagnetic
spin fluctuations \cite{mon}. However, experimental and
theoretical studies show that a gapless state can be induced in
\textit{s}-wave superconductors by residual interactions
\textit{coexisting} with the pairing interaction, such as magnetic
impurities \cite{abr}, spin density waves \cite{ove}, proximity
effects \cite{cla,pam} or phase fluctuations of the order
parameter \cite{cof,rod,eme}. In conclusion, a non-\textit{s}
symmetry implies a gapless state, but the opposite does not hold.

In this work we investigate whether the above predictions of
gapless state apply to the \textit{s}-wave
superconductor NbN. By measuring the variation of the magnetic
penetration depth $\lambda(T) - \lambda(0) \equiv \delta \lambda
(T)$ at $T\geq$ 1.5 K, we study the low energy excitations on
films with different microstructures obtained by varying growth
conditions. NbN is a model system for our purpose taking into
account the following: \textit{i}) the simple cubic structure with
fcc lattice \cite{bec}; \textit{ii}) the conventional BCS
\textit{s}-wave superconducting properties \cite{pam2};
\textit{iii}) the relatively high critical temperature $T_c
\approx 16$ K enables to study the low energy excitations in a
wider temperature region; \textit{iv}) various materials
preparation routes have been developed and optimized, especially
in view of electronics applications \cite{bac,vil1,vil2}.

For this work we have selected six NbN films prepared by DC
magnetron sputtering with similar $T_c$ close to the bulk value
$\approx$16 K but with different degrees of granularity obtained
by varying growth conditions. These conditions are reported in
Table I (see \cite{vil1, vil2} for more details): growth
temperatures are either room temperature or 620 $^\circ$C;
substrates are (1$\bar{1}$02) Al$_{2}$O$_{3}$ or (100) MgO single
crystals or (1$\bar{1}$02) Al$_{2}$O$_{3}$ single crystals with a
12 nm thick (100) MgO buffer layer; NbN thicknesses range from 10
to 1400 nm. Three different types of microstructure are
reproducibly obtained: \textit{i}) (100) epitaxial; \textit{ii})
highly (100) textured; \textit{iii}) (111) textured. Using (100)
MgO buffer layers, a high degree of (100) texture is obtained only
in sufficiently thin ($\textit{d} \lesssim$ 150 nm) NbN films
\cite{vil2}.

The film microstructure is characterized by X-ray diffraction in
the Bragg-Brentano geometry, X-ray pole figure measurements and
transmission electron and atomic force microscopies (TEM, AFM).
X-ray rocking curve analysis indicates that samples A1043 and
A1013 are epitaxial with FWHM of the (420) Bragg refelection
less than 1$^{\circ}$. The former sample is thin
(10 nm), whereas the latter is thick (950 nm). The film on the MgO
buffer layer (A1044) is thin (10 nm) and highly (100) textured.
The remaining three films on sapphire (A1057, A1060 and A1063) are
all thick (280-1400 nm) and display predominant (111) texture. The
different degrees of granularity in these films are confirmed by
AFM and TEM. For example, in sample A1043, we find an atomically
flat surface \cite{lam} and a crystal structure with apparently no
grain boundaries (see fig. 1a), which indicates that this film is
monocrystalline. A granular structure is evident in the (111)
textured sample A1060 from the TEM cross-sectional micrograph of
fig. 1b. We note a columnar structure with grains of size 15-35 nm
and well-defined grain boundaries, as previously reported for
films grown under the same conditions \cite{vil1}.

The microstructural properties of the six films are correlated
with DC resistivity $\varrho$ and DC transport critical current
density $j_c$ measurements. The results of these measurements are
summarized in Table I. In all films, $T_c$ is close to the bulk
value $\approx$16 K, which indicates that the microstructure does
not appreciably perturb the superconducting phase. Smaller
residual resistivities $\varrho$(20 K) $\approx$ 50$\mu\Omega$cm
and larger resistivity ratios $\varrho(300 \textrm{ K})/\varrho(20
\textrm{ K}) \approx$ 1.2 are found in the epitaxial films. The
highest values of $j_c$ (4.2 K) $\gtrsim$ 3 MA/cm$^2$ are found in
the epitaxial or thin (100) textured samples, while the three
thick (111) textured ones exhibit values $\lesssim$ 1 MA/cm$^2$
that decrease with increasing NbN thickness. In agreement with
earlier studies \cite{vil1} and with our TEM results, we conclude
the following: \textit{i}) the epitaxial samples have higher
crystalline quality; \textit{ii}) in the (111) textured samples,
the degree of granularity - defined as the degree of grain
decoupling - increases with increasing thickness.

To correlate the above structural and transport properties with
the low energy excitations, we have carried out $\delta\lambda(T)$
measurements in the range 1.5 K-$T_{c}$ using a single-coil mutual
inductance technique described elsewhere \cite{gau}. Typical
frequency and magnitude of the e.m. field perpendicular to the
film surface are 2-4 MHz and $<$ 0.1 mT respectively. Care has
been taken to always operate in the linear response regime, i.e.
below the Josephson critical field $B_{c1}$, monitored by varying
the applied magnetic field. Note that we measure an effective
magnetic penetration depth that, in the dirty limit, differs from
the London penetration depth by a correction factor of the order
of unity \cite{tink}. This limit is appropriate for our films,
for which we estimate a mean free path $\approx $1 nm and a coherence
length $\xi \approx 3-5$ nm \cite{vil1,lam}.
As a good approximation, the above
correction factor is temperature independent in the limited
temperature range of our measurements. Therefore the temperature
dependence of the variations $\delta\lambda(T)$ is not affected
by dirty limit effects.

In figs. 2a-b, we report the experimental $\delta \lambda (T)$
curves obtained. Only the epitaxial and thin (100) textured films
display the exponential behavior $\delta\lambda \sim
\textrm{exp}(-\Delta/{k_BT})$, characteristic of BCS
\textit{s}-wave superconductors \cite{bar} (see fig. 2a). All of
the three thick (111) textured samples exhibit a linear behavior
which can not be approximated by any power law (see fig.
2b); the range of linearity is limited below $\approx$ 2.7 K in
the thinner film A1057 and extends up to $\approx$4 K in the
thicker films A1060 and A1063. A BCS \textit{s}-wave fit perfectly
accounts for the experimental data of the three samples of fig.
2a. $\lambda(0)$ is the only free parameter of this fit, whereas
the ratio 2$\Delta(0)/k_BT_c$ is kept fixed at 4.6, taken from
tunneling data at 1.5 K on sample A1043. The fitted $\lambda$(0)
values are in the range 200-400 nm, in agreement with previous
reports \cite{pam2,vil2}. A BCS fit is satisfactory also for the
thinner (111) textured film A1057 of fig. 2b; indeed, the behavior
of this sample is intermediate between exponential and linear. For
the remaining two samples A1060 and A1063, no BCS fit is possible
for any plausible set of parameters.

These results give evidence of a gapless superconducting state in
the (111) textured films. We recall the mechanisms that may induce
a gapless state in a conventional BCS \textit{s}-wave
superconductor: \textit{i}) magnetic impurities; \textit{ii})
proximity effects; \textit{iii}) phase fluctuations of the order
parameter. In our case, the first two possibilities are excluded:
\textit{i}) the purity of the Nb target used is 99.995 \% and
traces of magnetic impurities are below 10 ppm. We
conclude that the amount of these impurities in the films is
negligible. In any case, the gapless samples contain no more
impurities than the others, since all of them are prepared with
the same Nb target; \textit{ii}) a previous TEM and tunneling
study \cite{vil1} indicates that the intergranular phases are
insulating, thus no proximity effects are expected. The third
picture remains plausible: as suggested earlier
\cite{deu,sim1,cof,rod,eme,cha}, both quantum and thermal phase
fluctuations can be relevant in granular superconductors with
short coherence length $\xi$ like our NbN films.

To verify the validity of the above picture in our case, we
consider an array of superconducting grains with intergranular
Josephson coupling energy $E_J$ and charging energy \textit{U}, as
first proposed by Abeles \cite{abe}, where \textit{U} is
associated with the excess of Cooper pairs \textit{n} in each
grain. \textit{U} and $E_J$ are related to the intergranular
capacitance \textit{C} and the critical current $I_{c}$
respectively, according to the relations $U=2e^{2}/C$ and
$E_{J}=\hbar I_{c}/2e$. Following previous work \cite{sim1,mcl},
we assume the array to form a square lattice of constant
\textit{a} equal to the average grain size. The resulting
Hamiltonian is:
\begin{eqnarray}
H=\frac{U}{2}\sum_{i} {n_i}^2 + E_{J}\sum_{i \neq j}
[1-\textrm{cos}(\vartheta_i-\vartheta_j)]
\end{eqnarray}
where $\vartheta_i$ is the phase of the order parameter in the
\textit{i}-th grain and $n_i=-2i\partial/\partial\vartheta_i$ is
its conjugate operator.

Eq. (1) describes both quantum and classical phase fluctuations.
Two arguments suggest a classical description of our results in
the range $T \geq$ 1.5 K of our experiment. \textit{i}) According
to theoretical studies \cite{cof,cha} supported by numerical
calculations \cite{rod}, quantum fluctuations lead to a reduction
of the renormalized superfluid density, implying a progressive
decrease of $\partial\lambda/\partial T$ as $T \rightarrow$ 0 K.
This effect is not observed in the curves of fig. 2b which
remain linear above 1.5 K. \textit{ii}) In our (111) textured
films, we estimate $U \ll E_J$, which corresponds to a purely
classical regime \cite{sim1,cof,cha}. Indeed, typical figures are:
$a \approx$ 20 nm, $\lambda (0) \approx $ 300 nm, $j_c \approx$
0.5-1 MA/cm$^2$, thickness and relative dielectric constant of the
intergranular regions respectively $t \lesssim $ 2 nm and
$\epsilon_r \gtrsim$ 2. Thus, we obtain $E_{J} \approx$ 50-100 meV
and $U \lesssim$ 10 $\mu$eV.

In the classical regime, the fluctuation strength is determined by
the ratio $E_J/k_BT_c$, where $T_c$ is the mean field critical
temperature. Using the above estimate, we find $E_J\gg
k_BT_c$ in our granular films. Thus, according to previous studies
\cite{sim1,mcl}, we expect small fluctuations which do not inhibit
long range phase order at any temperature. Indeed, we do not
observe any appreciable decrease of $T_c$ induced by granularity
($T_c$ is even slightly larger in the (111) textured films). These
fluctuations \textit{coexist} with the usual quasiparticle
excitations predicted for BCS \textit{s}-wave superconductors.
Both excitations lead to a progressive increase of $\lambda$ as
temperature increases. In the Gaussian approximation, valid at low
temperatures $T\ll E_J/k_B$ and in the absence of dissipation,
only long wavelength longitudinal fluctuations of arbitrarily
small energy are created (Goldstone modes). This produces a linear
increase $\delta \lambda \sim T$ \cite{cof,rod,eme} experimentally
observed in fig. 2b. On the other hand, the creation of
quasiparticles requires a finite thermal energy $\gtrsim\Delta$,
which leads to the exponential dependence of fig. 2a.

We now analyze in quantitative form the experimental curves of
fig. 2b within the framework of the above picture. Using Table I,
we estimate first the fluctuation contribution and then the
temperature at which the quasiparticle contribution becomes
significant. Following previous calculations \cite{rod,cof}, in
the Gaussian regime, our lattice model with \textit{U}=0 leads to:

\begin{eqnarray}
\frac{\partial \lambda}{\partial T} = \frac{ek_{B}}{\hbar}
\frac{1}{j_{c}a}
\end{eqnarray}
where $I_c \approx j_c a \lambda$ at 0 K and we neglect the weak
temperature dependence of this product at low temperature. In
Table I we report the values of $\partial \lambda /\partial T$
estimated using eq. (2) and compare them with the experimental
values. Taking into account the uncertainty in the estimate of
\textit{a}, we notice a good agreement. Interestingly, the trend
of $j_c$ in the (111) textured films follows the trend of
$\partial \lambda/\partial T$, as predicted by eq. (2): the
thicker the film, the smaller $j_c$ and the larger the slope
$\partial \lambda/\partial T$. In the films of fig. 2a,
exhibiting a BCS \textit{s}-wave behavior, $j_c$ is larger. Thus,
the fluctuation effect vanishes and only the quasiparticle effect
is observed. To estimate the temperature at which the two
effects become comparable, we separately evaluate their contribution to
$\partial\lambda/\partial T$ using Table
I and the BCS fit of fig. 2b. We find that the linear increase of
$\lambda$ produced by thermal phase fluctuations is expected to
dominate at $T \lesssim$ 4 K. This is in agreement with the point
of departure from linearity in samples A1060 and A1063.
As discussed before, in sample A1057 a gapless behavior is
visible but does not dominate yet above 1.5 K, since a BCS
fit is still satisfactory.

In conclusion, we have reported for the first time experimental evidence of gapless
state in a conventional BCS \textit{s}-wave superconductor, NbN.
Evidence is found in granular films exhibiting a clear linear dependence of
$\lambda$ in the range 1.5-4 K. Epitaxial or less granular films
exhibit the conventional dependence of \textit{s}-wave
superconductors and no trace of gapless state is found. A simple
model of granularity-induced thermal phase fluctuations of the
order parameter (Goldstone modes) quantitatively accounts for the above linear
dependence. No signature of quantum fluctuations is observed above
1.5 K, as expected taking into account the relatively large grains
of our films.

Our evidence of granularity-induced gapless superconductivity in
NbN raises the question whether this phenomenon is observable in
other granular superconductors. In cuprates, which display gapless
properties in agreement with a \textit{d}-wave symmetry model, it
should be investigated if these properties could be - at least
partially - explained by intrinsic granular properties associated
with chemical or electronic inhomogeneities in the coherence
length scale $\lesssim$ 3 nm. In any case, our results suggest
that the only reliable methods for testing non-\textit{s}
symmetries are those based on phase-sensitive measurements.

\begin{acknowledgments}
We acknowledge A. Andreone, A. Barone, B. K. Chakraverty and G.
Deutscher for stimulating discussions and A. Dubon for assistance
in the TEM work.
\end{acknowledgments}

$^{*}$Corresponding author. E-mail address:
gauzzi@maspec.bo.cnr.it

$^{\dag}$Present address is: INFM and Dipartimento di Scienze
Fisiche, Universit\`{a} di Napoli Federico II, 80125 Napoli,
Italy.

$^{\S}$Present address is: Laboratoire Pierre S\"{u}e, DRECAM,
CEA-Saclay, 91191 Gif sur Yvette, France.

\newpage

\begin{figure}
\caption{\textbf{a)}. Cross-sectional TEM image of the (100)
epitaxial sample A1043. Note the monocrystalline structure evidenced by
the contrast produced by the diagonal (110) planes. The
notation "R" indicates the (1$\bar{1}$02) orientation of the
sapphire substrate. \textbf{b).} The same as in \textbf{a)} for
the thick (111) textured A1060 sample. Note the granular structure
made of columnar grains.}
\end{figure}

\begin{figure}
\caption{\textbf{a).} Temperature dependence of the magnetic
penetration depth $\lambda$ measured in the films with
conventional BCS \textit{s}-wave behavior: A1013 (full circles),
A1043 (squares) and A1044 (open circles). \textbf{b)}. The same as
in \textbf{a)} for the films with gapless behavior: A1057
(lozenges), A1060 (full triangles) and A1063 (open triangles). The
BCS-like curve of sample A1044 (circles) of \textbf{a)} and the
corresponding BCS \textit{s}-wave fit (solid line) are also shown
for comparison.}
\end{figure}

\begin{table}
\label{table1} \caption{Summary of sample properties and
comparison between experimental values of $\partial \lambda /
\partial T$ and theoretical ones obtained using eq. (2). For the samples following the
BCS thermally activated behavior, $\lambda(0)$ is
obtained using a BCS \textit{s}-wave fit, as discussed in the
text.}
\begin{tabular}{|l|c|c|c|c|c|c|c|}
\hline
\textbf{sample}&{\textbf{A1043}}&{\textbf{A1013}}&{\textbf{A1044}}&{\textbf{A1057}}
&{\textbf{A1060}}&{\textbf{A1063}}\\
\hline \hline
\rm NbN thickness [nm]                 &   10          &   950         &    10         &   280         &   680          &    1400      \\
\rm substrate                          &Al$_{2}$O$_{3}$&MgO            &Al$_{2}$O$_{3}$/MgO & Al$_{2}$O$_{3}$ & Al$_{2}$O$_{3}$ & Al$_{2}$O$_{3}$ \\
\rm substrate orient.                  &(1$\bar{1}$02) &(100)    &(1$\bar{1}$02)/(100) & (1$\bar{1}$02)& (1$\bar{1}$02) & (1$\bar{1}$02) \\
\rm growth temp. [$^\circ$C]           &620            &room temp.     & 620           & room temp.    &room temp.      &room temp.    \\
\rm microstructure                     &epitaxial      &epitaxial      & highly textured  & textured      &textured        &textured      \\
\rm NbN orientation                    &(100)          &(100)          &(100)          &(111)          &(111)           &(111)      \\
\rm $T_{c}$ [K]                        &$14.2\pm0.1$   & $15.5\pm0.09$ & $14.4\pm0.1$  & $13.4\pm0.1$  &  $16.1\pm0.1$  &$16.3\pm0.1$   \\
\rm $j_{c}$ [MA/cm$^{2}$]              &$3.0\pm0.7$    &      -        & $9.0\pm0.3$   & $1.3\pm0.7$   &  $1.1\pm0.3$   &$0.64\pm0.04$  \\
\rm $\varrho$(20 K) [$\mu\Omega$\textrm{cm}]&    53    &      50       &         147   &      448      &  88            &  67           \\
\rm $\varrho$(300 K)/$\varrho$(20 K)   &          1.2  &      1.2      &          0.9  &      0.7      &  0.9           &  1.0         \\
\rm $\lambda(0)$ [nm] (BCS fit)        &240            &  260          & 270           & 470           &     -          &    -          \\
\rm $\partial\lambda/\partial T$ exp. [\AA/K] & 0      &      0        &      0        & $2.1\pm0.5  $ &  $3.3\pm0.2$   & $5.6\pm0.5$   \\
\rm $\partial\lambda/\partial T$ th.  [\AA/K] & 0      &      0        &      0        &      1.6      &  2.2           &        3.3    \\
\hline
\end{tabular}
\end{table}


\newpage
\begin{references}
\bibitem{gro}F. Gross, B. S. Chandrasekhar, D. Einzel, K. Andres, P. J. Hirschfeld, H. R. Ott, J. Beuers, Z. Fisk, and J.
L. Smith, Z. Phys. B: Condens. Matter \textbf{64}, 175 (1986).
\bibitem{har}W.N. Hardy, D.A. Bonn, D.C. Morgan, R. Liang, and K. Zhang, Phys. Rev. Lett. \textbf{70}, 3999 (1993).
\bibitem{bon}I. Bonalde, B.D. Yanoff, M.B. Salamon, D.J. Van Harlingen, E.M.E. Chia, Z.Q. Mao and Y. Maeno, Phys. Rev.
Lett. \textbf{85}, 4775 (2000).
\bibitem{mon}See, for example, P. Monthoux, A.V. Balatsky, and D. Pines, Phys. Rev. Lett. \textbf{72}, 1874 (1994) and references therein.
\bibitem{abr}A.A. Abrikosov and L.P. Gor'kov, Zh. Eksp. $\&$ Teor. Fiz. \textbf{39}, 178 (1960); P. Fulde, R. A.
Ferrel, Phys. Rev. \textbf{135}, A550 (1964); A.I. Larkin, Y.N.
Ovchinnikov, JETP \textbf{20}, 762 (1965).
\bibitem{ove}A.W. Overhauser and L.L. Daemen, Phys. Rev. Lett. \textbf{61}, 1885 (1988).
\bibitem{cla}J.H. Claassen, J.E. Evetts, R.E. Somekh, and Z.H. Barber, Phys. Rev. B \textbf{44}, 9605 (1991).
\bibitem{pam}M.S. Pambianchi, L. Chen and S.M. Anlage, Phys. Rev. B \textbf{54}, 3508 (1996).
\bibitem{cof}M.W. Coffey, Physica C \textbf{235-240}, 1961 (1995).
\bibitem{rod}E. Roddick and D. Stroud, Phys. Rev. Lett. \textbf{74}, 1430 (1995).
\bibitem{eme}V.J. Emery and S.A. Kivelson, Phys. Rev. Lett. \textbf{74}, 3253 (1995).
\bibitem{bec}K. Becker and F. Ebert, Z. f\"{u}r Physik \textbf{31}, 268 (1925).
\bibitem{pam2}M.S. Pambianchi, S.M. Anlage, E.S. Hellman, E.H. Hartford Jr., M. Bruns and S.Y.Lee, Appl. Phys. Lett. \textbf{64},
244 (1994); B. Komiyama, Z. Wang and M. Tonouchi, Appl. Phys.
Lett. \textbf{68}, 562 (1996).
\bibitem{bac}See, for example, D.D. Bacon, A.T. English, S. Nakahara, F.G. Peters, H. Schreiber, W.R. Sinclair, and R.B. van Dover, J. Appl.
Phys. \textbf{54}, 6509 (1983).
\bibitem{vil1}J.-C. Vill\'{e}gier, L. Vieux-Rochaz, M. Goniche, P. Renard,
M. Vabre, IEEE Trans Magn. \textbf{21}, 498 (1985); R. Chicault
and J.-C. Vill\'{e}gier, Phys. Rev. B. \textbf{36}, 5215 (1987).
\bibitem{vil2}J.-C. Vill\'{e}gier, B. Dela\"{e}t, V. Larrey, P. Febvre, J.W. Tao and G. Angenieux, Physica C \textbf{326-327}, 133 (1999).
\bibitem{lam}G. Lamura, PhD Thesis, Universit\'{e} Paris VI (2001).
\bibitem{gau}A. Gauzzi, J. Le Cochec, G. Lamura, B.J. J\"{o}nsson, V.A. Gasparov, F.R. Ladan, B. Pla\c{c}ais, P.-A. Probst, D. Pavuna and
J. Bok, Rev. Sci. Instr. \textbf{71}, 2147 (2000).
\bibitem{tink}M. Tinkham in \textit{Introduction to Superconductivity}
(McGraw-Hill, Singapore, 1996), p. 97.
\bibitem{bar}J. Bardeen, L. N. Cooper, and J. R. Schrieffer, Phys. Rev. \textbf{108}, 1175 (1957).
\bibitem{deu}G. Deutscher, Y. Imry and L. Gunther, Phys. Rev. B, \textbf{10}, 4598 (1974).
\bibitem{sim1}E. \u{S}im\'{a}nek, Solid state Comm. \textbf{31}, 419 (1979).
\bibitem{cha}B.K. Chakraverty, Physica C \textbf{341-348}, 75 (2000) and references therein.
\bibitem{abe}B. Abeles, Phys. Rev. B \textbf{15}, 2828 (1977).
\bibitem{mcl}W.L. McLean and M.J. Stephen, Phys. Rev. B \textbf{19}, 5925 (1979).
\end{references}
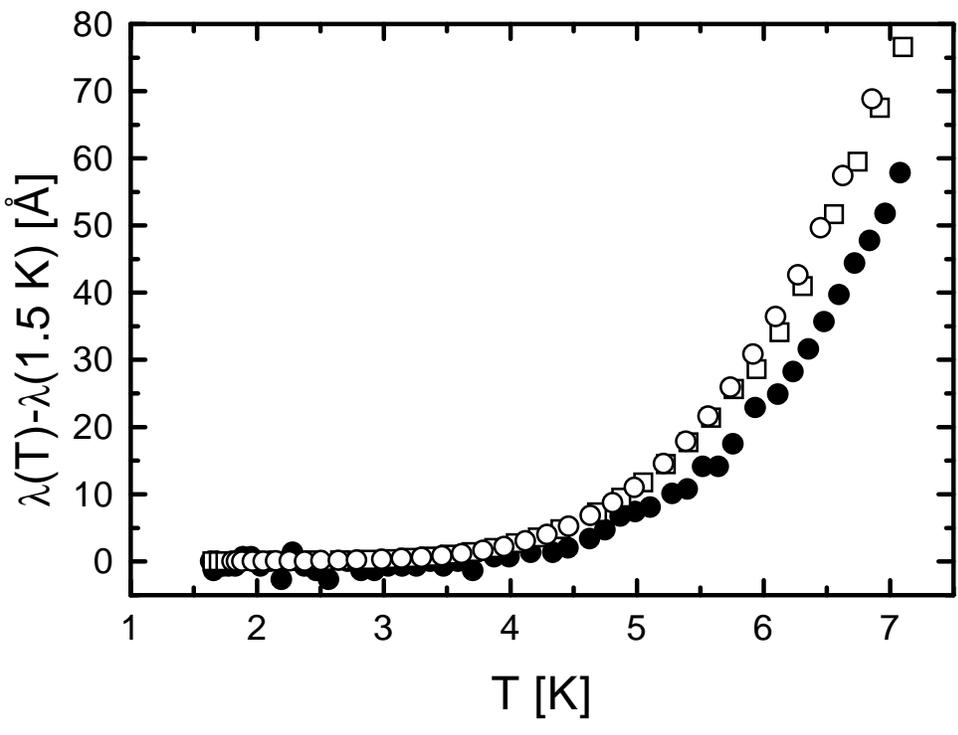
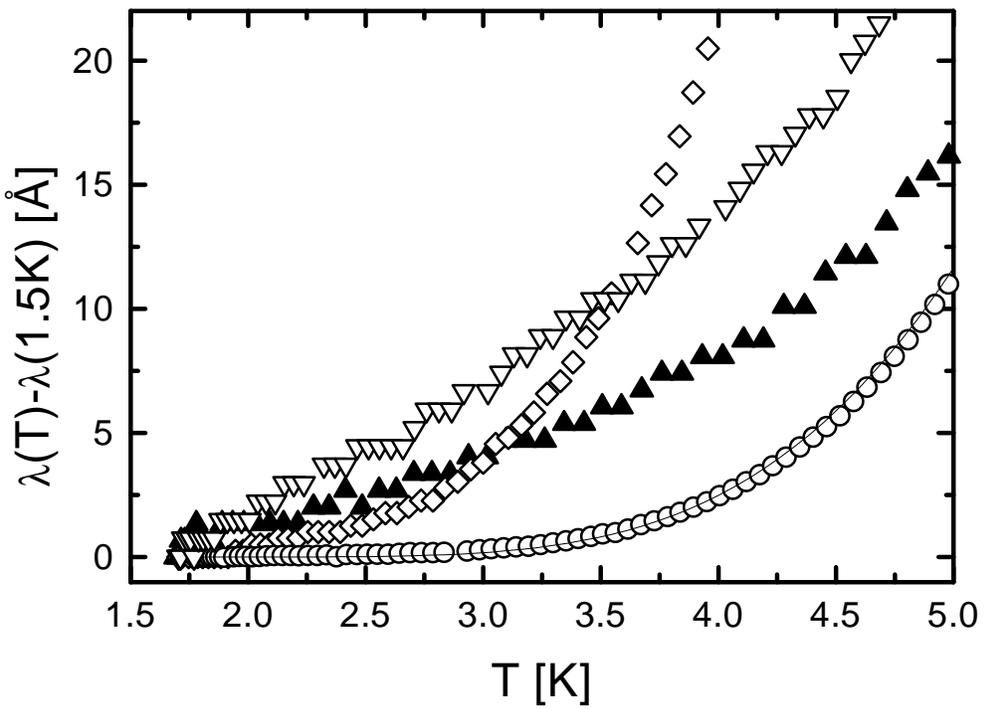
\end{document}